# An Initial Assessment of a Clear Air Turbulence Forecasting Product

Ankita Nagirimadugu

Thomas Jefferson High School for Science and Technology

Alexandria, VA

**Abstract**

Clear air turbulence, also known as CAT, can cause damage to an aircraft's structure and, in severe cases, harm passengers. Though CAT has been thoroughly studied since the mid 1960's, scientists have not been able to create an accurate forecasting device. The product tested is known as the Deformation-Vertical Shear Index (DVSI), created by Knox, Ellrod, and Williams. The DVSI is currently used by several commercial carriers and military aircraft. The general feedback has been positive; however, results indicate that the product tends to overestimate CAT intensity.

**Introduction**

Clear air turbulence (CAT) can be seen on visible satellite images near transverse cirrus bands. CAT is caused by various atmospheric factors including pressure, jet stream location, mountain waves, cold and warm fronts, and nearby thunderstorms. In addition, pilots and forecasters have confirmed that CAT can occur near mountain ranges when affected by these conditions (Knox, Ellrod, and WIlliams, 2006). In certain cases however, CAT was detected in clear skies, puzzling both scientists and pilots alike. To better understand the production of CAT, scientists have conducted numerous studies on fluid dynamics.

Generally, CAT is caused by eddies, or changes in wind flow direction that cause distrubences. Kelvin–Helmholtz Instability (KHI) plays a key role in CAT creation. In terms of fluid dynamics, the friction caused by two media moving at different speeds is similar to the creation of CAT in the atmosphere. According to the Encyclopedia of Atmospheric Sciences, “KHI episodes are the cause of a large fraction of CAT.” Furthermore, wind-shear overcoming the stability causes turbulence intensification and eddy formation (Holton, Curry, and Pyle, 2002).

There are several aspects to keep in mind while discussing KHI and CAT. Turbulence does not merely exist as a condition, but as a process. KHI usually lasts for 5 minutes and the intensity depends on the shear layer. CAT occurs in a “surge-like” fashion, forecasting even more difficult (Holoton, Curry, and Pyle, 2002).

Along with KHI, another producer of CAT is the Internal Gravity Wave (IGW). IGW is another form of waves that develops in the atmosphere. The waves are caused by gravity acting on density in the atmosphere. If IGW has a large amplitude, then an aircraft flying through will suffer strong, periodical gusts of wind. These cause KHI, which then lead to CAT. Conclusively, CAT intensity depends on KHI as well as IGW (Holoton, Curry, and Pyle, 2002).

Because CAT is a potential threat to flight safety, its observation began as early as the 1960s by the means of pilot reports (PIREPS). In fact, some governments require commercial and personal aircrafts to be equipped with a predictive wind-shear radar. The first PIREPS were collected through the International Civil Aviation Organization. However, an important factor to keep in mind is that PIREPS are highly subjective and depend on the pilot experience, as well the size, structure, and the weight of the airplane.

Through careful analysis of a large variety of PIREPS, it was concluded that CAT correlates with the jet stream (Knox, Ellrod, and WIlliams, 2006).

Common methods of predicting CAT include calculating the horizontal wind-shear, the vertical wind-shear, or the Richardson Number. The DVSI, also known as the turbulence index (TI), is the product of resultant deformation (A) and vertical wind-shear (B).

$$\mathrm{DVSI} = \underbrace{[(\delta u/\delta x - \delta v/\delta y)^2 + (\delta v/\delta x + \delta u/\delta y)^2]^{½}}_{A} \underbrace{(\delta V/\delta z)}_{B} \quad (1)$$

The DVSI was created with the data from the North American Model (NAM) and the Rapid Update Cycle (RUC2). The use and testing of this product occurred through the National Oceanic and Atmospheric Association (Knox, Ellrod, & WIlliams, 2006).

**Methodology**

The validation of this process was broken down into three parts. First, using the National Oceanic and Atmospheric Administration's (NOAA) website: ftp://www.orbit.nesdis.noaa.gov/pub/smcd/opdb/aviation/turb/rucverif the PIREPS were compiled into a database. The information recorded in the database included the DVSI, PIREP, location, and the time of the image recording. The product images were taken every three hours and were presented in universal time (UTC). Second, the images were analyzed by highlighting the location of the jet stream, relatively high DVSI predictions, and questionable or uncertain points. The jet stream was labeled using daily weather maps available at: http://www.hpc.ncep. noaa.gov. The wind direction and speed shown on the maps is at the 500 millibar level where as the jet stream is typically found near the 300 millibar level indication that the jet stream location was approximated. A questionable point was defined as a point that had a PIREP that was found to be smaller than the predicted DVSI. Finally, in order to validate the questionable point, radar and satellite imagery from the Plymouth State University website was used to determine whether or not the plotted data point was in the presence of a convective storm. The radar and satellite images were available hourly. By highlighting the location on all three images it was to determine whether the storm was affecting the data. If there was a storm present in the radius, and the associated radar reflectivity was above 50dbz, the PIREP was removed from the database. The point was also removed if the intensity was lower than 50dbz and covered a vast area. All the questionable data points were analyzed and compiled into a separate database.

**Case 1: July 5, 2007, 0Z**

Figure 1 represents the image produced using the DVSI. The red line is the approximate location of the jet stream. The highlighted locations in yellow are those with high predicted values. The PIREPS are presented in bold blue on the image, and the combination of letters and numbers below it is the aircraft type. The orange circle represents a questionable region that could possibly be affected by a convective storm. The same criteria hold true for all product images. It should be noted that high predicted values correspond with the location of the jet stream, bolstering the observation that CAT has a tendency to occur near the jet stream.

Figure 2 is a radar image acquired from Plymouth State University website. The image is based on intensities. Any intensity above 50dbz is considered strong enough to remove the questionable point. As shown, this image displays the need for the removal of the questionable point due to the existence of a storm in the region that could have possibly skewed the data.

Figure 3 is a satellite image from Plymouth State University website. This once again confirms presences of a storm in the Illinois region.

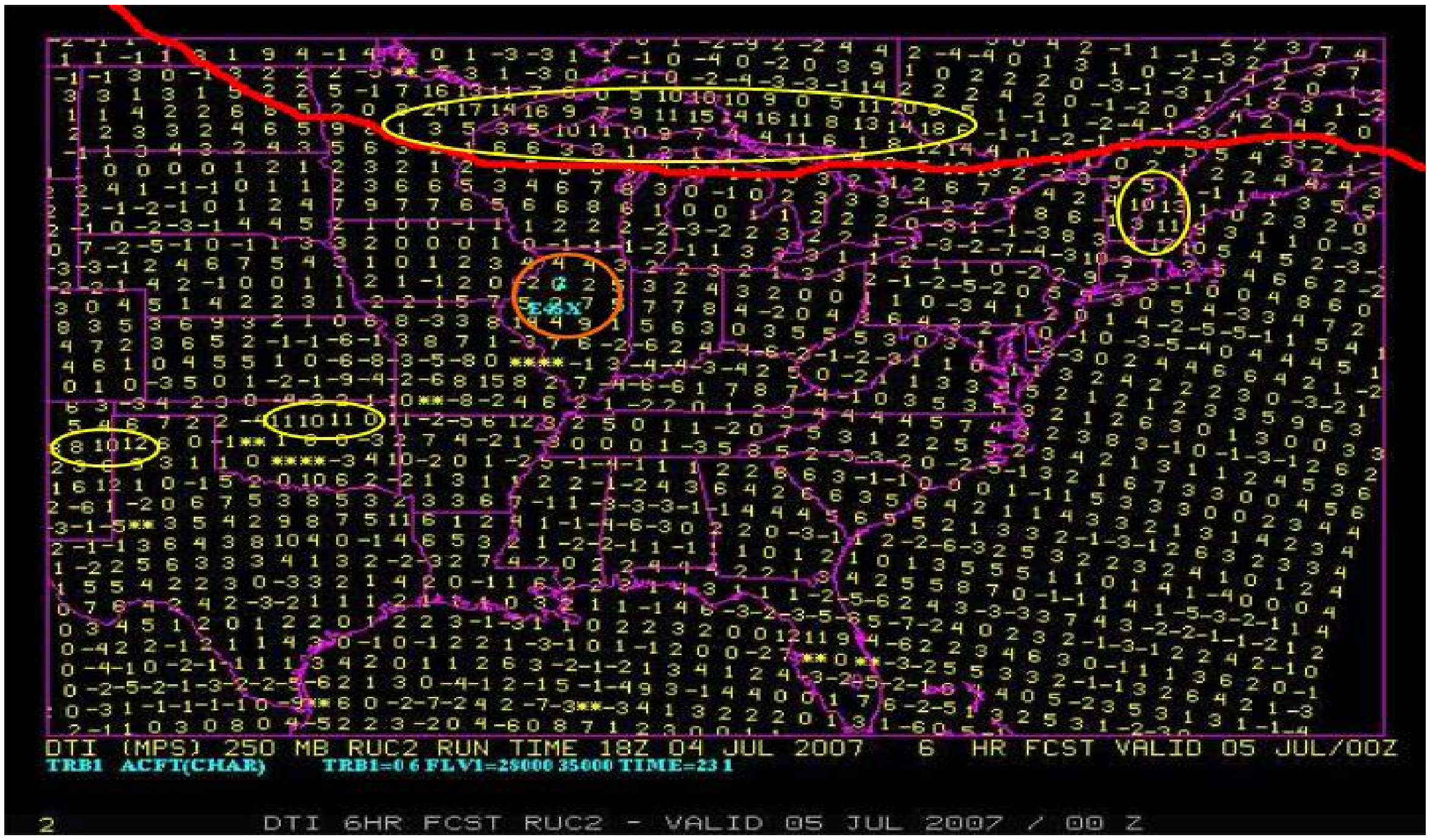

**Fig.1: Image of forecasted CAT, point in Illinois removed**

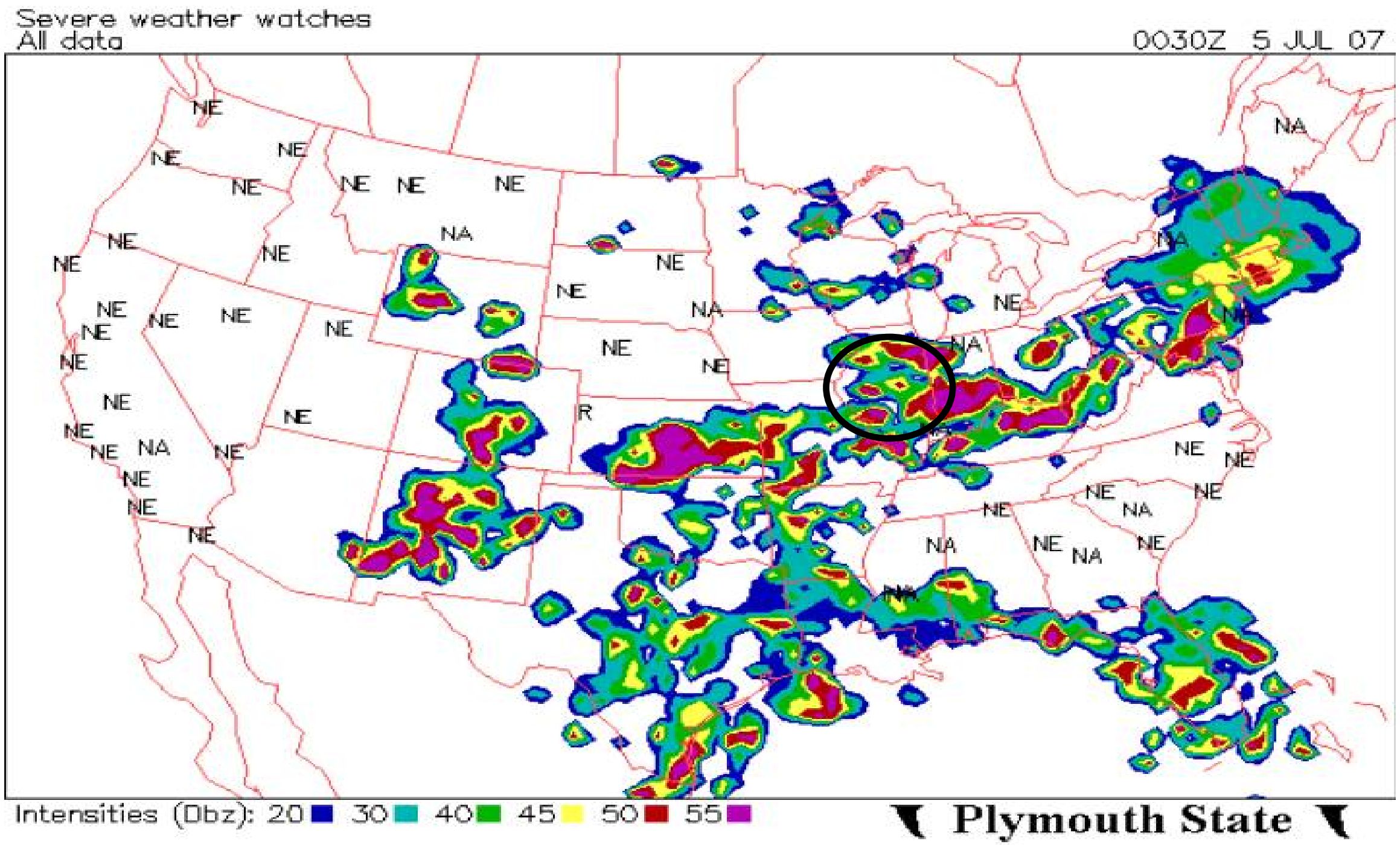

**Fig.2: Radar image, doubtful point in Illinois removed due to storm in the indicated radius**

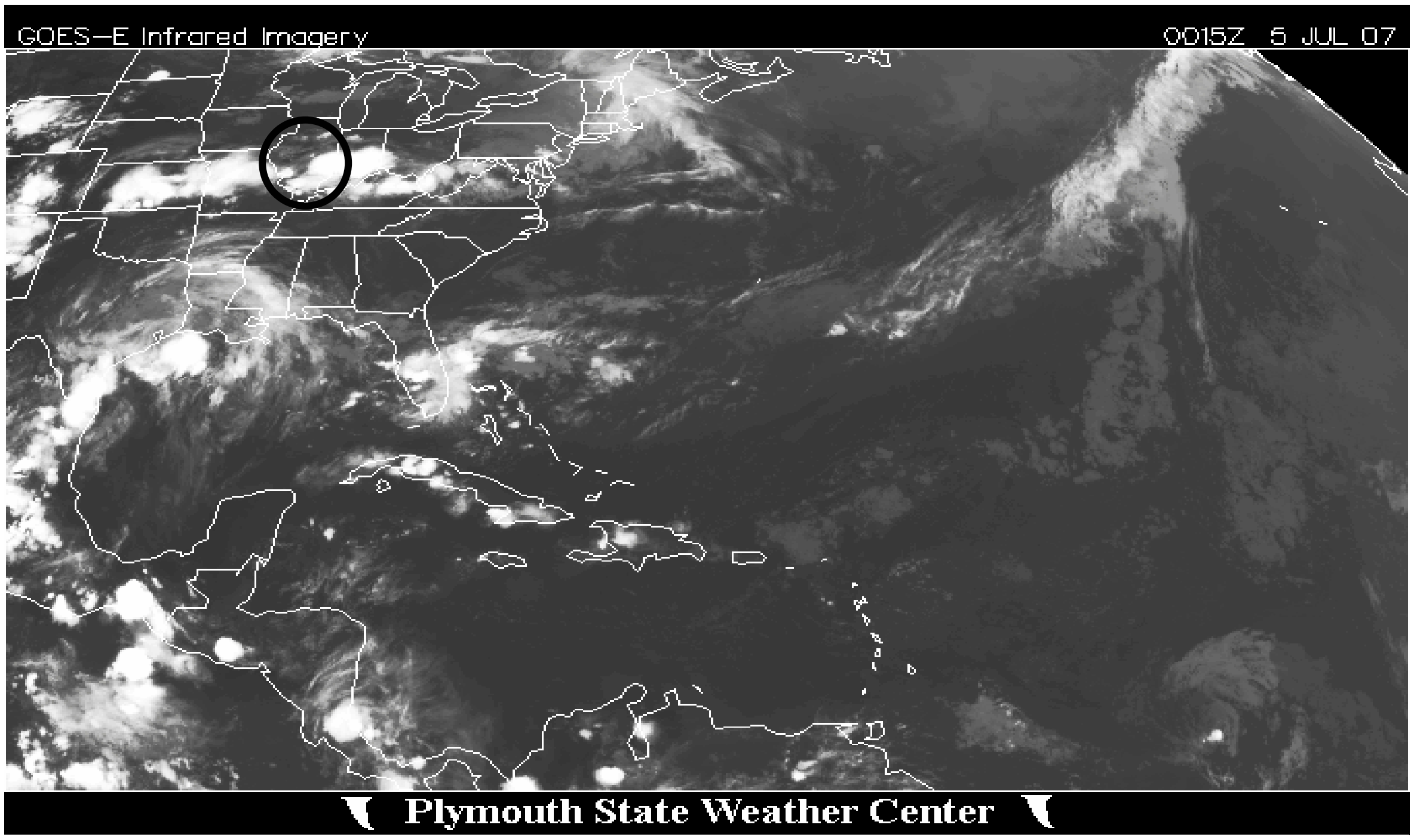

**Fig. 3 Satellite Image, doubtful point in Illinois removed due to storm in the indicated radius**

## Case 2: July 5, 2007, 15Z

Figure 4 once again reinforces the observation that CAT occurs near the jet stream. These product images show that the predicted value is significantly lower than the real time data. When compared to Figure 5 it is easy to see that there are no storms present that could skew the data. Figure 6 shows thin clouds in the region that are clearly visible. These clouds appear to be cirrus clouds due to their thinness and darkness. As mentioned before, CAT can be produced around cirrus bands, especially near the jet stream.

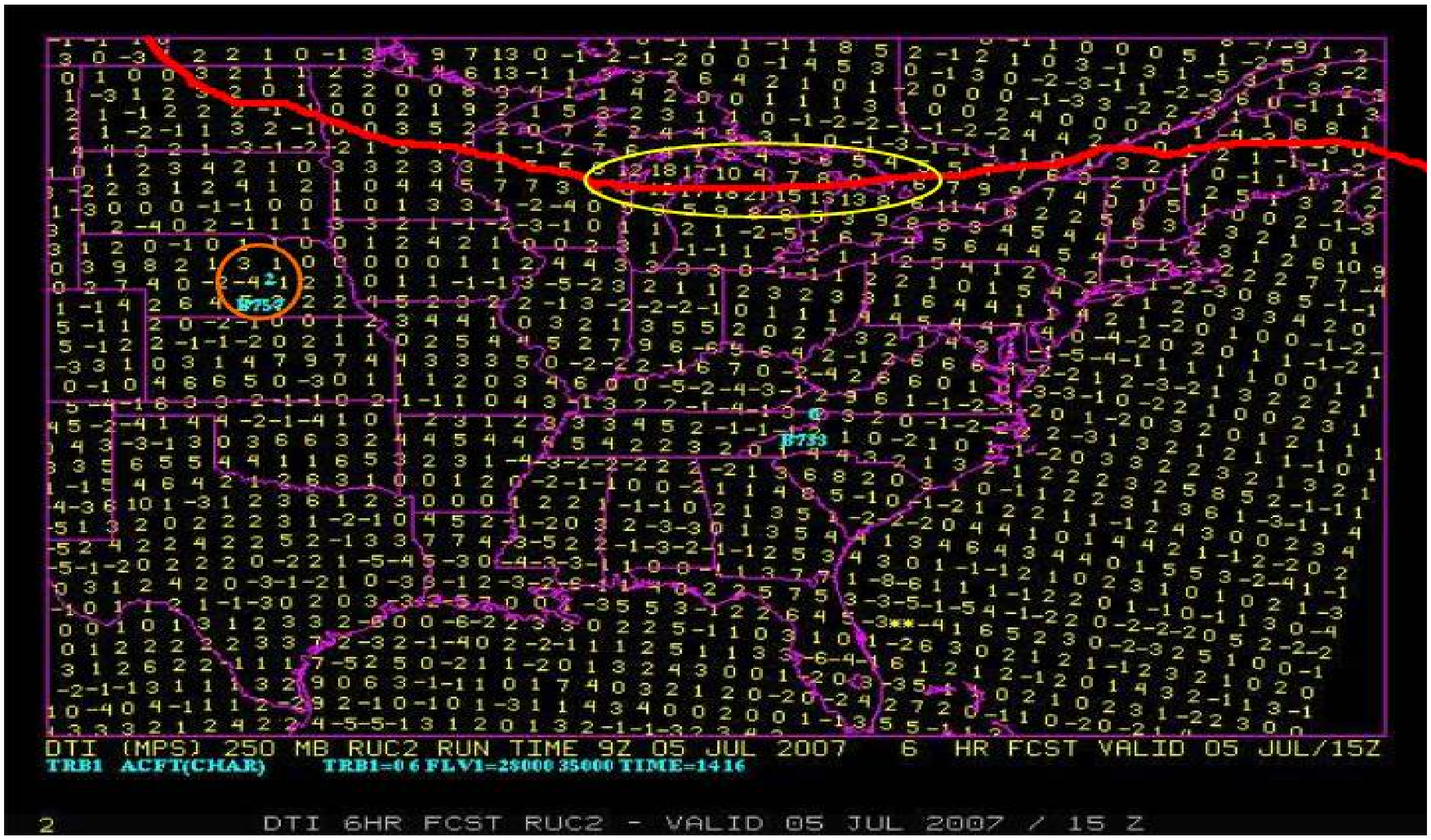

**Fig.4: Image of forecasted CAT, point in Nebraska not removed**

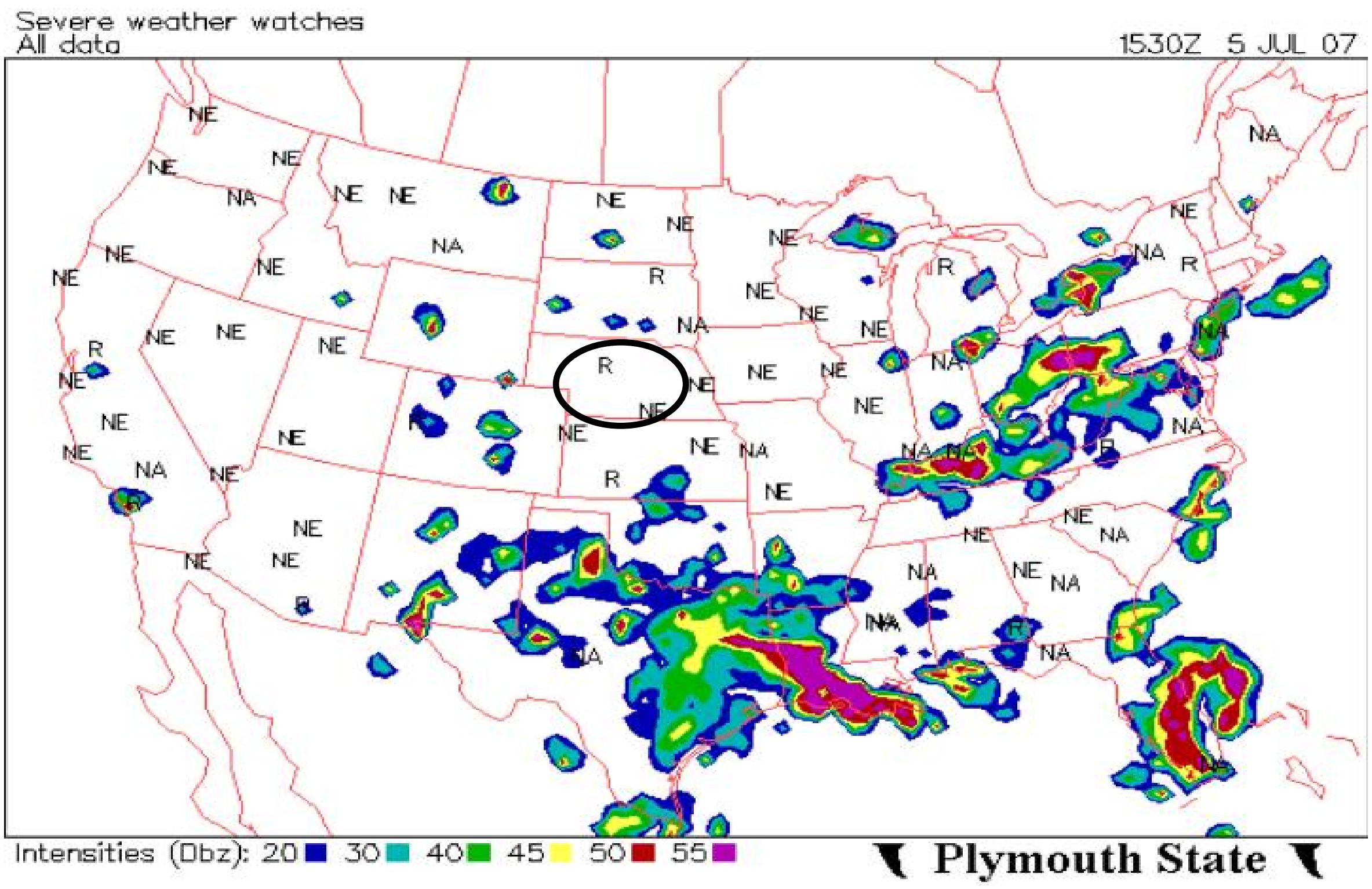

**Fig.5: Radar image, doubtful point in Nebraska is not affected by the storms**

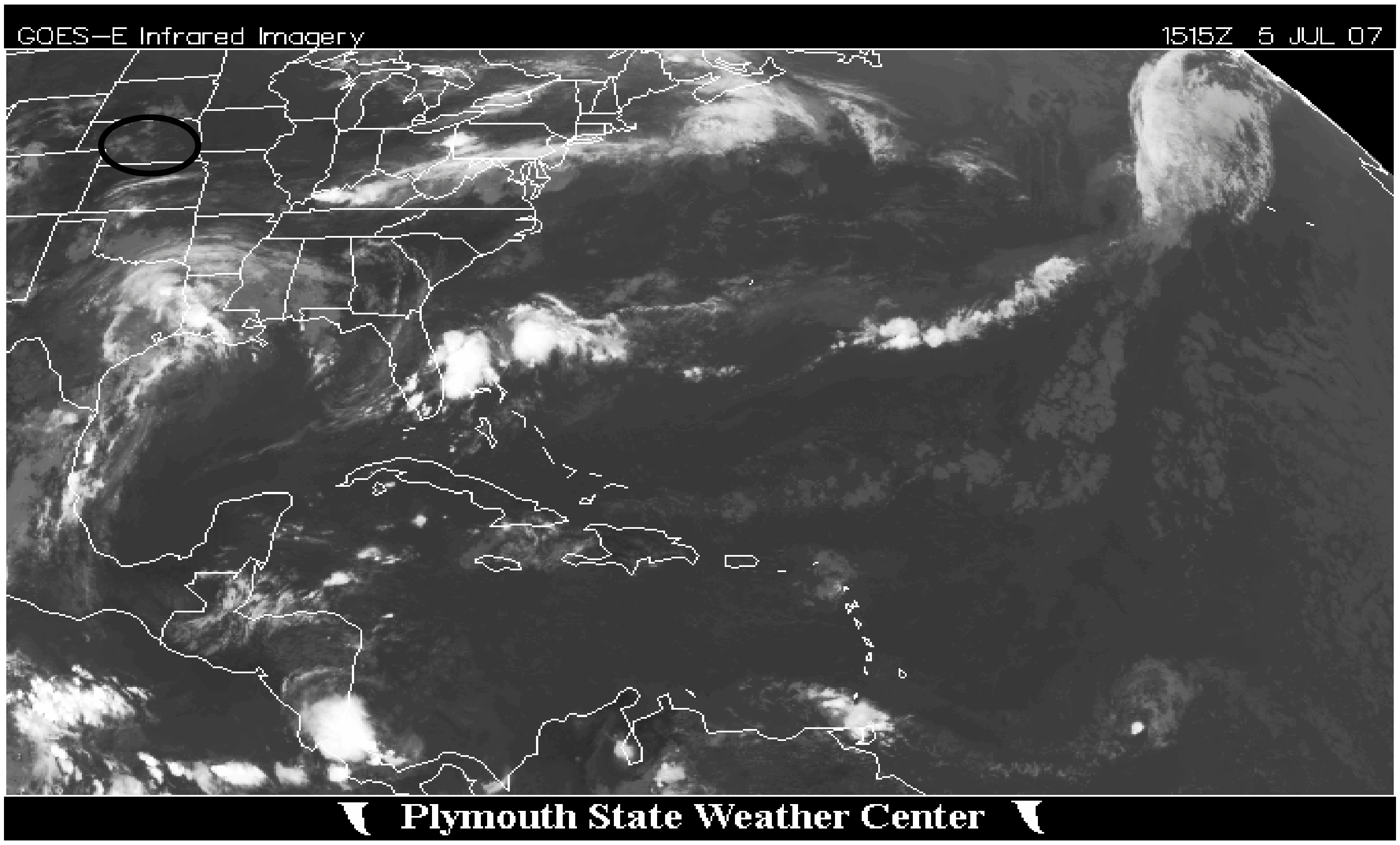

**Fig. 6 Satellite Image, doubtful point in Nebraska not removed**

**Case 3: July 3, 2007 21Z**

The following images illustrate the product working effectively. The predicted values correspond with the actual values as seen in Figure 7. One can see on radar and satellite images that the product performed accurately when there were no storms present. The product supplied not only one accurate point but three, in different regions confirming that it wasn't by chance.

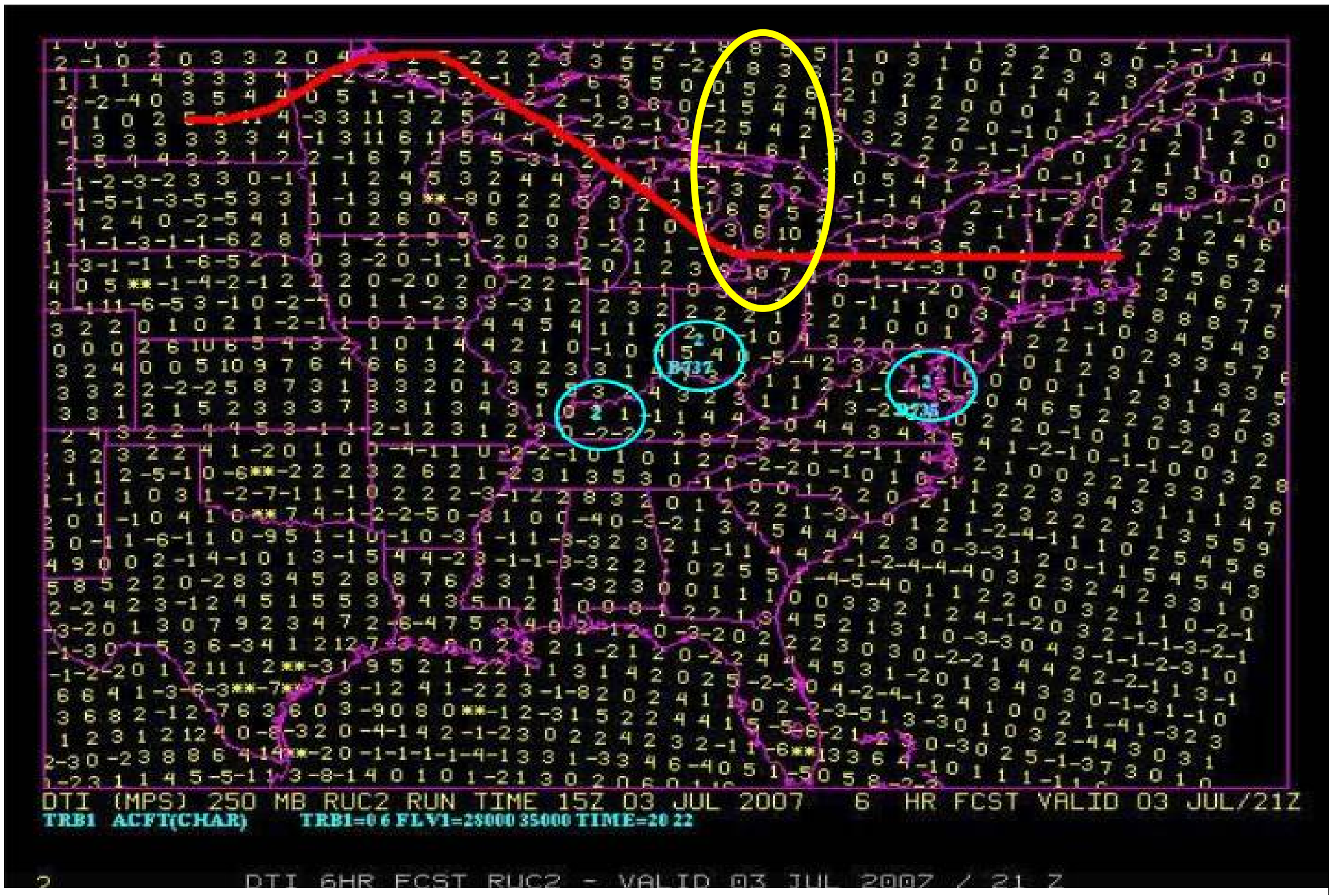

**Fig. 7 Image of forecasted CAT, points in Kentucky, Ohio, and Maryland**

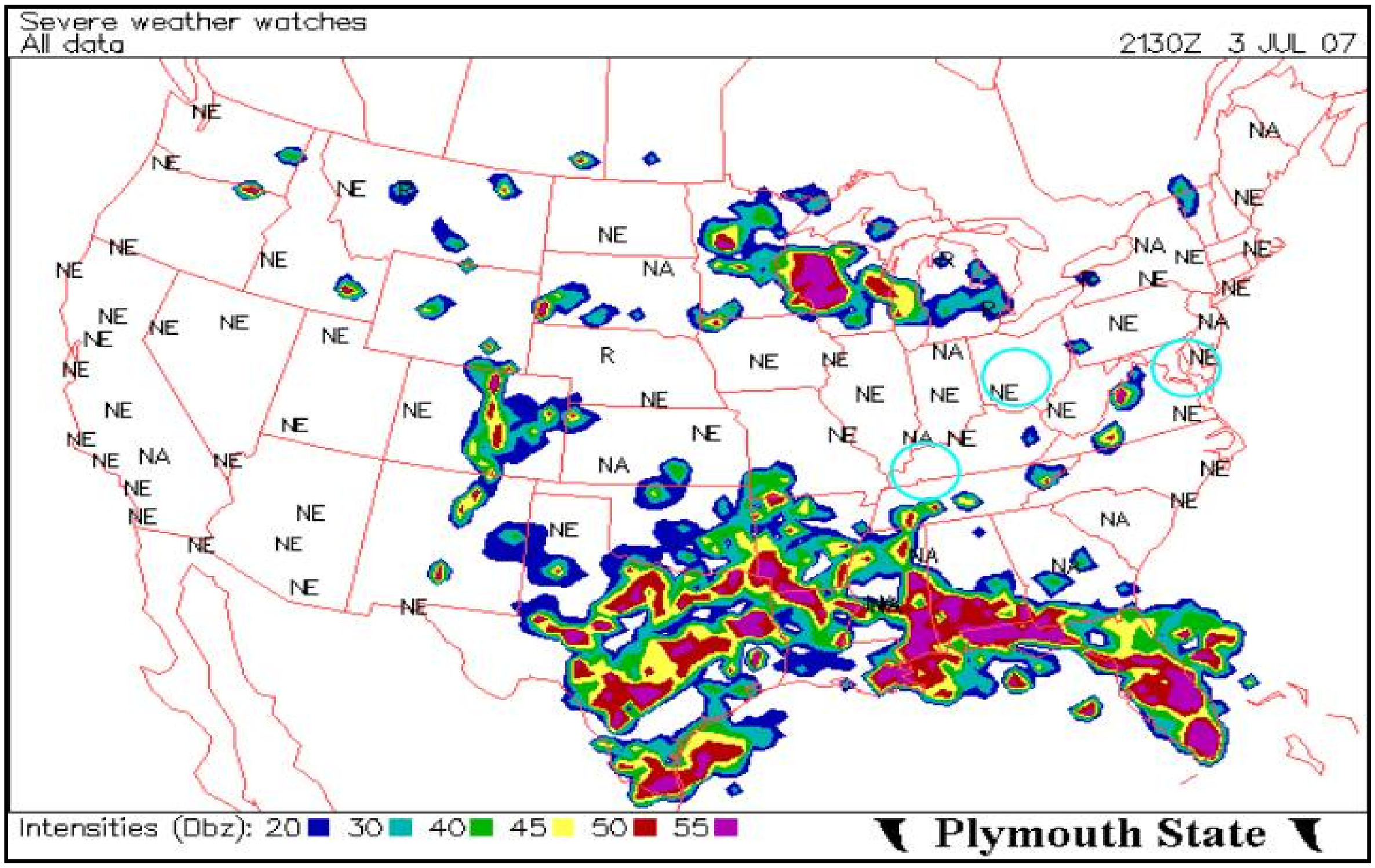

**Fig. 8 Radar image of forecasted CAT, points in Kentucky, Ohio, and Maryland**

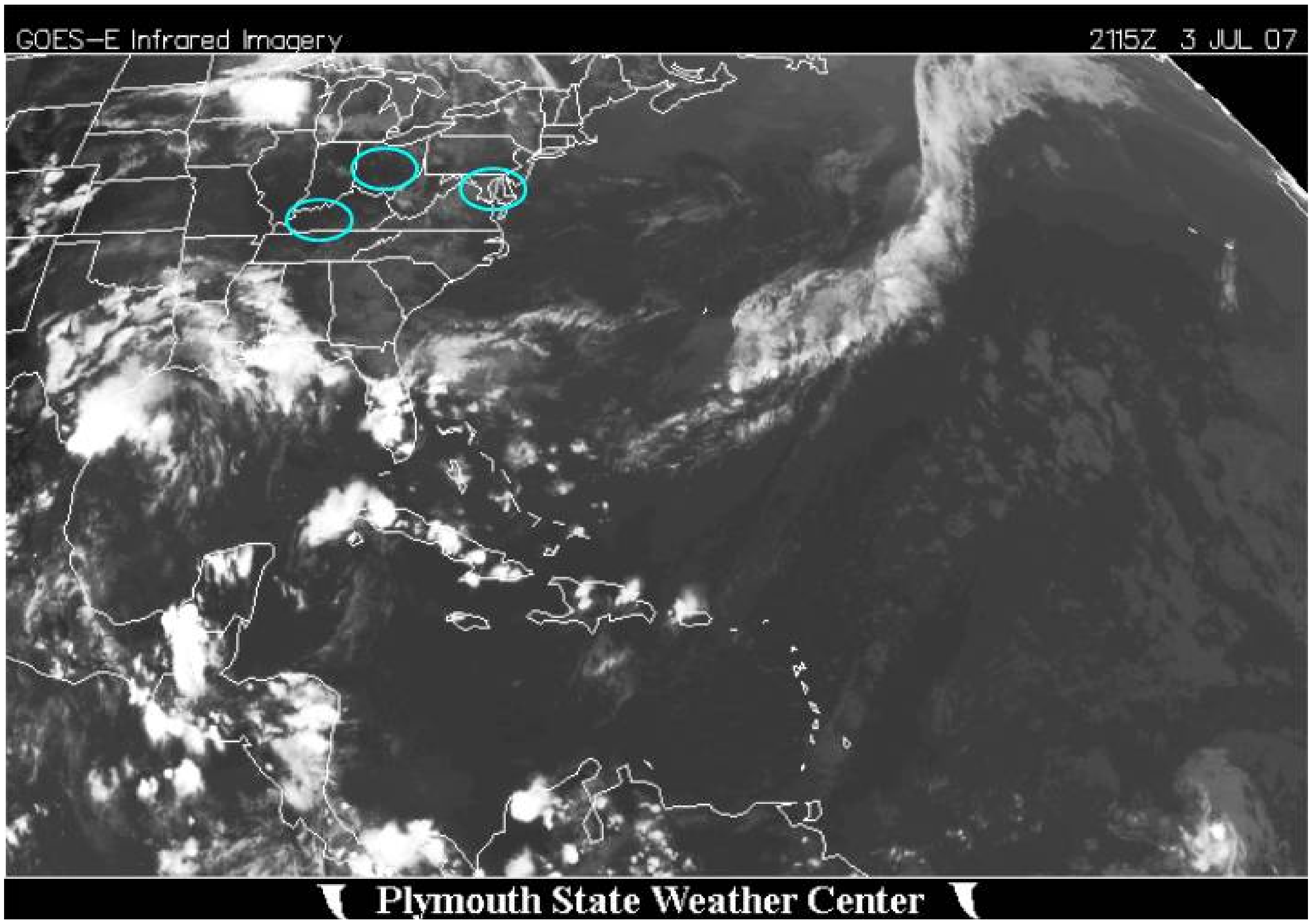

**Fig. 9 Satellite image of forecasted CAT, points in Kentucky, Ohio, and Maryland**

**Graph of DVSI compared to PIREPS**

After the questionable points were removed, the remaining points were plotted. When the DVSI was compared to the PIREPS it resulted in a very low correlation rate of 0.0197 possibly because of the PIREPS highly subjective nature.

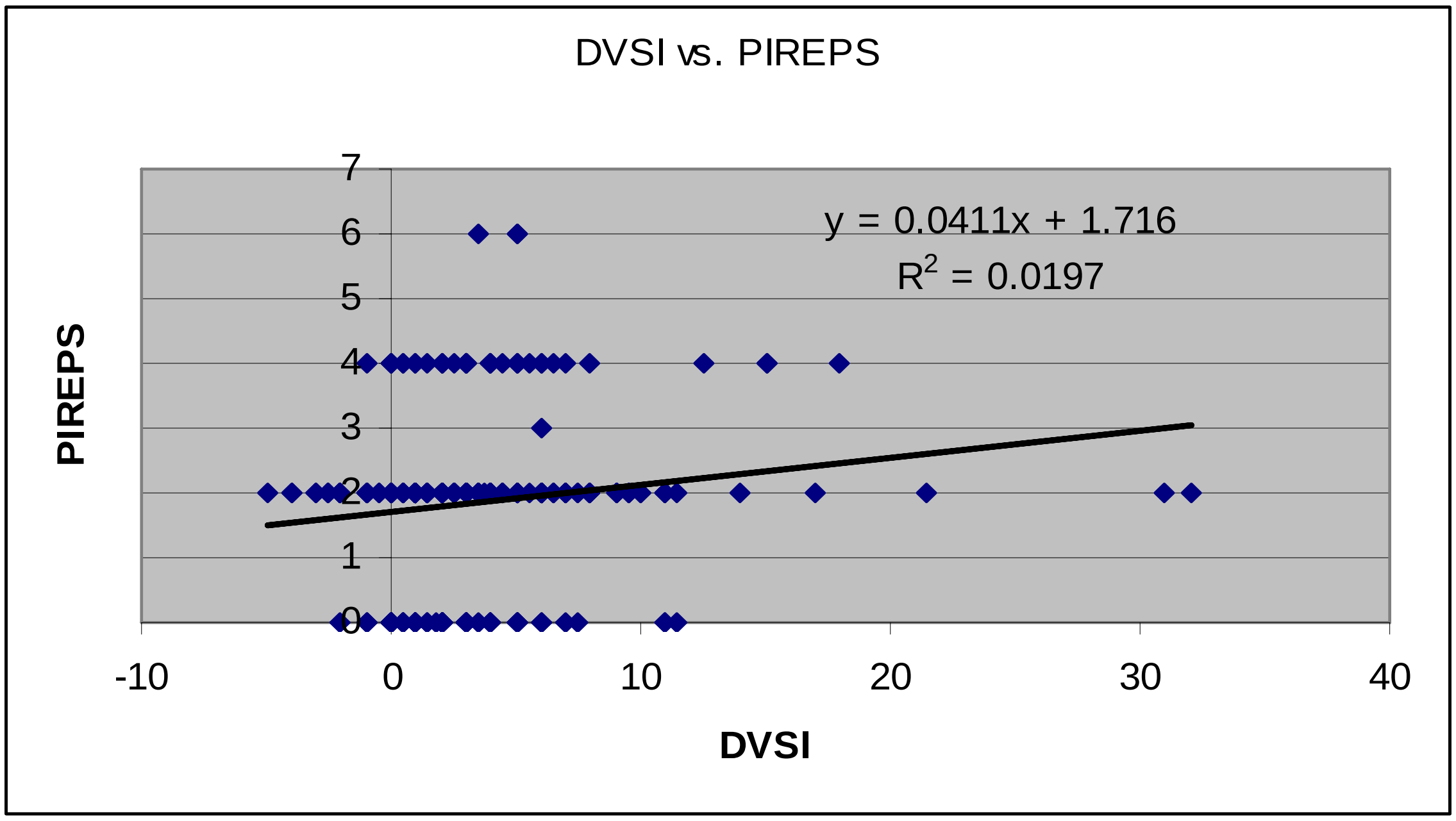

**Fig. 10 The correlation between the DVSI and PIREPS after the removal of doubtful points**

**Conclusion**

The three cases above and the scatterplot in Figure 10, clearly indicate that this product requires further testing before being considered reliable. All three cases hold true of the fact that high predicted values occur near the jet stream. In addition, the second case demonstrated a situation in which the DVSI underestimated turbulence intensity, evidenced by the PIREP of significant turbulence in a region indicating low DVSI values. When the satellite image was observed, it was shown that particular areas might be prone to CAT due to cirrus clouds in the region. With the use of all three images, pilots maybe able to avoid CAT.

CAT is caused by various other factors that do not play a role in the DVSI algorithm. Those factor maybe the reason for the low correlation value. Perhaps further testing on IGW and KHI will help prevent unreasonably high DVSI values. The product does not underestimate very often, but it does overestimate, causing pilots to take unnecessary avoidances.

As mentioned earlier, the airlines that use the DVSI are content with its forecasting abilities and utilize their own methods of interpreting the DVSI. Dissimilar types of aircrafts sense turbulence differently, hence, data was split up by aircraft types. Because of this, the correlation of particular aircrafts was not significantly higher than the combined data. It showed that the type of aircraft did not play a role in the low correlation value. Overall, it is vital to understand that this is only a preliminary assessment. The findings presented today show that there needs to be further testing in order to acquire a conclusive evaluation of this product.